# Electrical spin injection from an organic-based ferrimagnet in a hybrid organic/inorganic heterostructure

Lei Fang[1*], K. Deniz Bozdag[1*], Chia-Yi Chen[2], P.A. Truitt[1], A. J. Epstein[1,3†] and E. Johnston-Halperin[1†]

[1] Department of Physics, The Ohio State University, Columbus, OH 43210-1117, USA
[2] Chemical Physics Program, The Ohio State University, Columbus, OH, 43210-1106, USA
[3] Department of Chemistry, The Ohio State University, Columbus, OH 43210-1173, USA

Abstract

We report the successful extraction of spin polarized current from the organic-based room temperature ferrimagnetic semiconductor V[TCNE]$_x$ ($x$~2, TCNE: tetracyanoethylene; $T_C$ ~ 400 K, $E_G$ ~ 0.5 eV, $\sigma$ ~ $10^{-2}$ S/cm) and its subsequent injection into a GaAs/AlGaAs light-emitting diode (LED). The spin current tracks the magnetization of V[TCNE]$_{x~2}$, is weakly temperature dependent, and exhibits heavy hole / light hole asymmetry. This result has implications for room temperature spintronics and the use of inorganic materials to probe spin physics in organic and molecular systems.

The field of semiconductor spintronics promises the extension of spin-based electronics beyond memory and magnetic sensing into active electronic components with implications for next-generation computing [1] and quantum information [2]. The development of organic-based magnets with room temperature magnetic ordering [3] and semiconducting functionality [4] promises to further broaden this impact by providing a route to all-organic spintronic devices [5-6] and hybrid organic/inorganic structures capable of exploiting the multifunctionality [7-8] and ease of production in organic systems [9] as well as the well established spintronic functionality of inorganic materials. Our work demonstrates electrical spin injection in a hybrid organic/inorganic spin-resolved light-emitting diode (spin-LED) [10] structure and opens the door to a new class of active, hybrid spintronic devices with multifunctional behavior defined by the optical, electronic and chemical sensitivity [11] of the organic layer. In addition, spin transport in these hybrid structures provides the opportunity to use well-characterized inorganic materials as a probe of spin physics in organic and molecular systems.

The magnetic order in V[TCNE]$_{x\sim2}$ ($T_C \sim$ 400 K) originates from direct antiferromagnetic exchange coupling between the unpaired spins of V$^{2+}$ ($t_{2g}$, $S = 3/2$) and TCNE$^-$ ($\pi^*$, $S = 1/2$). As shown in Fig. 1(a), the unpaired spin in the TCNE$^-$ anion is distributed over the entire molecule [12] and occupies the $\pi^*$ antibonding level. The $\pi^*$ orbital can accept another electron with opposite spin, which costs an additional Coulomb energy [4, 13] $U_c \sim$ 2 eV. Therefore, the $\pi^*$ band is split into two oppositely spin polarized subbands: occupied $\pi^*$ and unoccupied $\pi^*+U_c$. The $t_{2g}$ levels lie within the Coulomb gap [14] and define the valence band while the $\pi^* + U_c$ levels define the conduction band with an 0.5 eV bandgap ($\sigma_{300K} \sim 10^{-2}$ S/cm). This proposed electronic structure with ferromagnetically aligned conduction and valance bands is consistent with theoretical calculations [15-17] and experimental studies [5, 14, 18].

The implications of the electronic and magnetic structure of V[TCNE]$_{x\sim2}$ for a spin-LED device are depicted schematically in Fig. 1(b). Spin polarized carriers are extracted from a spin-polarized source (V[TCNE]$_{x\sim2}$ here) and injected into the conduction band of the $n$-AlGaAs layer of an $n$-$i$-$p$ diode with a GaAs quantum well (QW) embedded in the intrinsic region. The sign and magnitude of the spin-polarized charge injection are determined by the magnetization of the V[TCNE]$_{x\sim2}$ and can be analyzed through the polarization of the heavy-hole (HH) and light-hole (LH) electroluminescence (EL) from carriers that relax into the quantum well. These optical polarizations have complementary coupling to the spin polarization of the electron current [19], where the optical polarization is defined as $P_{EL} = (I^{RCP} - I^{LCP})/(I^{RCP} + I^{LCP})$ ($I^{RCP}$ and $I^{LCP}$ indicate the intensity of right and left circular polarization, respectively). For example, an $S = + 1/2$ electron will yield an RCP photon on recombining with a HH and an LCP photon on recombining with a LH. This asymmetry of HH and LH with a given magnetization of the spin polarized source is shown schematically in Fig. 1(c) and can be resolved in EL due to the energy difference between HH and LH states. Thus the optical polarization of the EL is directly proportional the spin polarized current present in the V[TCNE]$_{x\sim2}$.

The organic and inorganic layers of the spin-LED are synthesized using chemical vapor deposition (CVD) and molecular beam epitaxy (MBE), respectively. The AlGaAs/GaAs heterostructure follows previous spin-LED devices. The MBE grown III-V material is grown on a $p$-doped ($p = 1\times10^{18}$ cm$^{-3}$) GaAs (100) substrate with layer structure as follows: 300 nm $p$-GaAs ($1\times10^{17}$ cm$^{-3}$)/ 200 nm $p$-Al$_{0.1}$Ga$_{0.9}$As ($p = 1\times10^{16}$ cm$^{-3}$)/ 25 nm $i$-Al$_{0.1}$Ga$_{0.9}$As/ 10 nm $i$-GaAs/ 25 nm $i$-Al$_{0.1}$Ga$_{0.9}$As/ 100 nm $n$-Al$_{0.1}$Ga$_{0.9}$As (n = $1\times10^{16}$ cm$^{-3}$ )/ 15 nm $n\rightarrow n^+$Al$_{0.1}$Ga$_{0.9}$As, $n = 1\times10^{16}$ cm$^{-3}$ to $5\times10^{18}$ cm$^{-3}$/ 15 nm $n^+$-Al$_{0.1}$Ga$_{0.9}$As ($n^+ = 5\times10^{18}$ cm$^{-3}$) [20]. Devices are fabricated using standard photolithography to define a mesa that is protected with photoresist except for a narrow window (100 μm by 1 mm). The V[TCNE]$_{x\sim2}$ is deposited uniformly across the sample at 40 ℃ and 35 Torr in an argon glovebox [9]. The top electrical contact consists of an optically transparent aluminium layer (7 nm) and a high conductivity gold layer (23 nm, Fig. 2(b) inset). Due to the air-sensitivity of CVD-prepared V[TCNE]$_{x\sim2}$ the samples are transferred within the glovebox to a custom-designed air-free sample mount for transfer to a magneto-optical cryostat. See Auxiliary Material 1a for additional details.

The electrical and optical properties of both V[TCNE]$_{x\sim2}$ spin-LEDs and bare LEDs are studied and compared (Fig. 2(a); Auxiliary Material 1b). Both devices show modified *p-i-n* diode behaviour with the V[TCNE]$_{x\sim2}$ spin-LED showing an additional series impedance consistent with thermally activated transport in an intrinsic semiconductor [4] with a bandgap of 0.5 eV (Auxiliary Material 2a). Moreover a positive linear magnetoresistance is observed in V[TCNE]$_{x\sim2}$ spin-LED devices with the same sign and magnitude as isolated V[TCNE]$_{x\sim2}$ films [4], confirming charge flow through the V[TCNE]$_{x\sim2}$ layer (Auxiliary Material 2b). Figure 2(b) shows an EL spectrum from a V[TCNE]$_{x\sim2}$ spin-LED at a temperature of 60 K and a bias of +18.5 V collected using a 0.3 m spectrometer and LN$_2$ cooled charge-coupled device (CCD). The peak at 1.533 eV represents the transition from the ground state of the conduction band of the quantum well to the HH states of the valence band. The higher energy peak at 1.541 eV is attributed to the corresponding LH transition, consistent with theoretical predictions for the HH/LH splitting [21]. The broad peak centred at 1.471 eV is due to recombination in the p-doped GaAs substrate.

The EL polarization, $P_{EL}$, of the HH and LH transitions is measured and analysed as a function of the out of plane magnetic field at $T = 60$ K, with $I = 0.5$ mA and $V = +18.5$ V. At each field a series of spectra with alternating helicity are acquired and analyzed using a variable wave plate and a linear polarizer in the collection path [10].

$I^{RCP}$ and $I^{LCP}$ are determined individually by integrating over the appropriate spectral peaks (highlighted regions in Fig. 2(b)) and $P_{EL}$ is calculated for the HH and LH, respectively. In these spin-LED devices the total polarization signal is composed of four independent contributions: the spin injection EL, the intrinsic magnetic field response of the AlGaAs/GaAs quantum well (Zeeman effect) [22], magnetic circular dichroism (MCD) from the cryostat windows, and MCD [10, 23-24] in the V[TCNE]$_{x\sim2}$ layer. Figures 3 (a) – (c) show cartoons explaining how these contributions combine to give the total measured EL polarization while Figures 3 (d) – (f) show corresponding experimental measurements of these effects.

Figure 3 (a) shows the expected linear response from both the Zeeman effect and dichroism in the cryostat windows. In principle the HH should show a three times larger Zeeman splitting than the LH due to their increased angular momentum ($m_j = 3/2$ and $m_j = 1/2$, respectively); however this difference is not observed (Fig. 3 (d)), suggesting that the dichroism of the cryostat windows is the dominant effect.

In Fig. 3 (b) the expected MCD response is plotted as the dashed gray line, with the contribution from Fig. 3 (a) plotted as the dashed orange line. The total of these two effects is shown as the green solid line, and qualitatively agrees with independent measurements of the MCD shown in Fig. 3 (e). This data is collected by *optically* exciting unpolarized carriers in the quantum well using a linearly polarized pump at 1.771 eV and a power density of 100 W/cm$^2$. The resulting photoluminescence (PL) exhibits both the linear window dichroism shown in Fig. 3(d) and saturation at $\pm 200$ Oe, consistent with the field dependence of the magnetization of V[TCNE]$_{x\sim2}$ (Fig. 4(a), green line; Auxiliary Material 3) as predicted in Fig. 3 (b). Note that the since the HH and

LH PL are so close in energy, there is no measureable difference between the HH and LH MCD response.

Finally in Fig. 3 (c), the spin injection EL is added to the various background signals shown in Figs. 3 (a) and (b) (green line). In contrast to these backgrounds, the spin injection EL should have opposite sign for the HH and LH (red and blue dashed lines, respectively). The total EL signal is given by $P_{EL}^{HH} = P_{spin}^{HH} + P_{MCD}^{HH} + P_{Zeeman+window}^{HH}$ and $P_{EL}^{LH} = P_{spin}^{LH} + P_{MCD}^{LH} + P_{Zeeman+window}^{LH}$ for the HH and LH, respectively. Note that in the case where the magnitude of the spin injection signal is comparable to that of the MCD background the two terms that depend on the magnetization of the V[TCNE]$_{x\sim2}$ cancel for the LH, leaving a simple linear dependence due to Zeeman splitting and window dichroism. Figure 3 (f) shows the total EL signal from a full V[TCNE]$_{x\sim2}$ spin-LED device. As expected, and in contrast to the background measurements in Figs. 3 (d) and (e), a clear distinction between HH and LH polarization is observed and the linear response of the LH indicates that at this temperature the spin injection signal and MCD are of comparable magnitude.

The bare EL signal ($P_{bare}$) for both HH and LH transitions can be removed by a linear fit to the calibration data in Fig. 3(d), leaving the corrected polarization as $P_{EL}^{HH/LH} = P_{MCD}^{HH/LH} + P_{spin}^{HH/LH}$. The background corrected polarization signals are plotted in Fig. 4(a) where it can be seen that $P_{EL}$ of the HH peak reaches saturation at ~ 200 Oe, closely tracking the out of plane magnetization of the V[TCNE]$_{x\sim2}$ (Fig. 4(a) green line). Moreover, subtracting $P_{EL}^{LH}$ from $P_{EL}^{HH}$ will result in a cancellation of the MCD terms and give twice the optical polarization ($2 \times P_{spin}$). This analysis yields a saturated spin polarization of $P_{spin}^{sat}$ =0.098 ± 0.007% (the error is one standard deviation at saturation). The saturated spin polarization is related to the injected electron spin polarization at the quantum well ($\Pi_{inj}$) by $P_{spin}^{sat} = \frac{\Pi_{inj}}{\eta}$, where $\eta = (1+\frac{\tau_r}{\tau_s})$ and $\tau_r$ and $\tau_s$ are the recombination time and spin relaxation time of the injected electrons, respectively [19]. Independent photoluminescence studies give an $\eta$ of 8.48 (Auxiliary Material 4), and thus a $\Pi_{inj}$ of 0.83 ± 0.07% at 60 K. This number is comparable to the initial results from GaMnAs and Fe based spin-LED structures [10, 22].

The dependence of $P_{EL}$ on bias and temperature is explored in Figs. 4(b) and 4(c), respectively. The large impedance of the V[TCNE]$_{x\sim 2}$ layer, and the resulting high turn on voltage for the spin-LED, limits the range of applied bias over which measureable EL can be obtained. Figure 4(b) shows a measurement of $P_{EL}$ at a bias of +15.2 V and a current of 0.08 mA (the minimum values that give measureable luminescence), showing the same $P_{EL}$ as at +18.5 V (Fig. 4a) to within the resolution of the measurement. This insensitivity to bias is initially surprising given that the 3.3 V difference between measurements is roughly twice the bandgap of the III-V materials. However, comparison with a bare LED device (Fig. 2 (b)) suggests that the voltage drop across the *p-i-n* junction is only 0.27 V and it is important to note that this measurement probes the ground state of the quantum well and thus is sensitive only to carriers that have fully thermalized (see Auxiliary Material 5 for a full discussion).

Figure 4(c) shows $P_{EL}$ at a temperature of 140 K, the highest temperature at which EL from the quantum well can be clearly resolved. This data provides further support for the simple two-component model for the polarization presented above. At 140K the magnitude for both $P_{spin}^{HH}$ and $P_{spin}^{LH}$ are reduced from 60K, consistent with an increased $\eta$ of 18.09 (see Auxiliary Material 4) and consistent with literature reports of the decrease in spin relaxation time at high temperature [25]. When this $\eta$ is combined with $P_{spin}^{sat} = 0.036 \pm 0.006\%$, a value for $\Pi_{inj}$ (140 K) of 0.66 $\pm$ 0.16% is determined. This relatively modest dependence on temperature ($\Pi_{inj}$(60 K) = 0.83 $\pm$ 0.07%) is consistent with spin injection from other room temperature ferromagnets such as Fe [24][26][27].

In conclusion, optical detection of electrical spin injection across a V[TCNE]$_{x\sim2}$/AlGaAs interface has been demonstrated using an active hybrid organic/inorganic spin-LED device. We report circular polarization of the electroluminescence that tracks the magnetization of the V[TCNE]$_{x\sim2}$ layer and exhibits HH/LH asymmetry. These studies validate the spintronic functionality of organic-based magnets, laying the foundation for a new class of multifunctional hybrid spintronic structures and representing the first all-semiconductor spintronic device with the potential for room temperature operation. Moreover, we establish the technique of using extensively studied inorganic heterostructures as a sensitive probe of free carrier spin physics in organic and molecular systems.

We acknowledge Dr. Mark Brenner and Semiconductor Epitaxial and Analysis Laboratory (SEAL) of the Ohio State University for providing AlGaAs/GaAs quantum well LED detector. We also acknowledge discussions and interactions with J. A. Gupta, R. Myers and J. -W. Yoo. This work was supported by the NSF MRSEC program (DMR-0820414), OSU-Institute of Material Research and DOE – spin (DE-FG02-01ER45931).

# References

* These authors contributed equally to this work

†To whom correspondence should be addressed. E-mail: epstein@mps.ohio-state.edu and ejh@mps.ohio-state.edu.

[1] International technology roadmap for semiconductors. Executive summary for 2009. (2009).
[2] D. P. Divincenzo, Science **270**, 255 (1995).
[3] J. M. Manriquez *et al.*, Science **252**, 1415 (1991).
[4] V. N. Prigodin *et al.*, Adv. Mater. **14**, 1230 (2002).
[5] J.-W. Yoo *et al.,* Nat. Mater. **9**, 638 (2010).
[6] J.-W. Yoo *et al.,* Phys. Rev. B. **80**, 205207 (2009).
[7] J.-W. Yoo *et al.,* Phys. Rev. Lett. **97**, 247205 (2006).
[8] K. Deniz Bozdag *et al*., Phys. Rev. B. **82**, 094449 (2010)
[9] K. I. Pokhodnya, A. J. Epstein, and J. S. Miller, Adv. Mater. **12**, 410 (2000).
[10] Y. Ohno *et al.,* Nature **402**, 790 (1999).
[11] J. S. Miller and A. J. Epstein, Angew. Chem. Int. Edit. **33**, 385 (1994).
[12] A. Zheludev *et al.,* J. Am. Chem. Soc. **116,** 7243 (1994).
[13] C. Tengstedt *et al.,* Phys. Rev. B. **69,** 165208 (2004).
[14] C. Tengstedt *et al.*, Phys. Rev. Lett. **96,** 057209 (2006).
[15] A. L. Tchougreeff and R. Dronskowski, J. Comput. Chem. **29,** 2220 (2008).
[16] H. Matsuura, K. Miyake, and H. Fukuyama, J. Phys. Soc. Jpn. **79,** 034712 (2010).
[17] G. C. De Fusco *et al.*, Phys. Rev. B. **79**, 085201 (2009).
[18] J. B. Kortright *et al.*, Phys. Rev. Lett. **100**, 257204 (2008).
[19] F. Meier and B. P. Zakharchenya, *Optical Orientation* Amsterdam, The Netherlands: North-Holland, vol. 8. (1984).
[20] C. Adelmann *et al.*, Phys. Rev. B. **71,** 121301(R) (2005)
[21] B. Jonker, Proc. IEEE. **91,** 727 (2003).
[22] H. Zhu *et al.,* Phys. Rev. Lett. **87,** 016601 (2001).
[23] X. Y, Dong *et al.*, Appl. Phys. Lett. **86,** 102107 (2005).
[24] A. T. Hanbicki *et al.*, Appl. Phys. Lett. **80,** 1240 (2002).
[25] J. M. Kikkawa and D. D. Awschalom, Phys. Rev. Lett. **80**, 4313 (1998)
[26] The spin injection efficiency tracks with the bulk magnetization, and is distinct from materials such as LSMO [27] where surface relaxation leads to a rapid depolarization of the spin current at elevated temperature.
[27] J.-H. Park *et al.*, Phys. Rev. Lett. **81,** 1953 (1998)

FIG. 1 (color online). (a) Schematic energy diagram for V[TCNE]$_{x\sim2}$. (b) Schematic band diagram of the spin-LED. The study of the V[TCNE]$_{x\sim2}$/$n$-AlGaAs interface through which spin-polarized electrons are injected is complicated by the large and nonlinear resistance from bulk V[TCNE]$_{x\sim2}$, which is represented by the dashed box. (c) The expected polarization signal for HH/LH transitions (right) for a given magnetization of the spin injector (left). Due to quantum selection rules, HH and LH show opposite polarization behavior.

FIG. 2 (color online). (a) $I$–$V$ curves for both bare LED (triangles) and V[TCNE]$_{x\sim2}$ spin-LED (diamonds) devices at $T = 60$ K. (b) Typical EL spectrum of a V[TCNE]$_{x\sim2}$ spin-LED device ($I = 0.5$ mA, $V = +18.5$ V) at $T = 60$ K. The shaded areas in the spectrum indicate the region of polarization integration over the quantum well heavy-hole (HH) and light-hole (LH) peaks, respectively. Inset: A schematic of the V[TCNE]$_{x\sim2}$ spin-LED device structure.

FIG. 3 (color online). (a) - (c) Cartoon showing the evolution of the detected luminescence signal from three different sources. (a) Linear background from the intrinsic magnetic field response of the AlGaAs/GaAs quantum well and dichroism from the cryostat windows. (b) Linear background from (a) plus the magnetic circular dichorism (MCD) response from the magnetic layer (c) Expected HH (red) and LH (blue) EL from spin injection and backgrounds from (a and b). (d)-(e) Experimental results corresponding to cartoons (a) – (c), EL from a non-magnetic device, MCD from photoluminescence and EL from a spin-LED, respectively. Filled symbols are for field sweeping up and open symbols are for field sweeping down. Note that a field-independent offset is subtracted from all scans. Error bars are determined using the standard deviation of multiple measurements at each field value.

FIG. 4(color online).Circular polarization for HH and LH transitions in a V[TCNE]$_{x\sim2}$ spin-LED device at (a) $T = 60$ K, $I = 0.5$ mA and $V = +18.5$ V, (b) $T = 60$ K, $I = 0.08$ mA and $V = +15.2$ V and (c) $T = 140$ K, $I = 1.5$ mA and $V = +10.1$ V (red and blue symbols for HH and LH, respectively). The solid green line in (a) represents out of plane magnetization of the V[TCNE]$_{x\sim2}$ layer. Error bars are determined as in Fig. 3 except for (b), where the long measurement time required for low bias measurements prevents statistically significant averaging. As a result, the error is estimated from the scatter of the LH data.

Fig. 1.

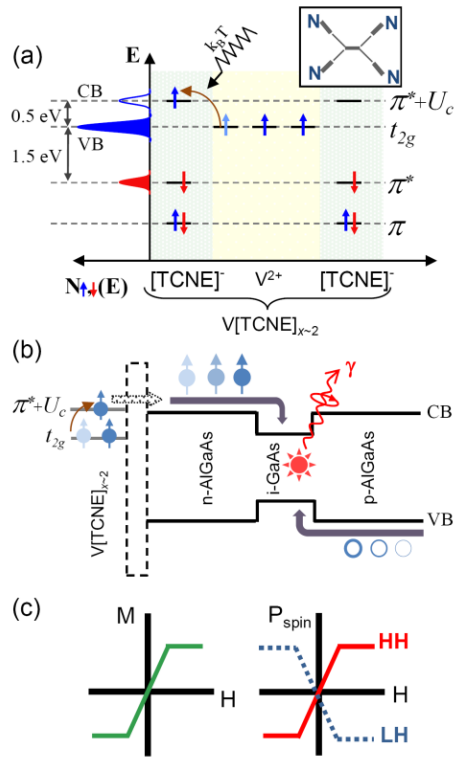

Fig. 2

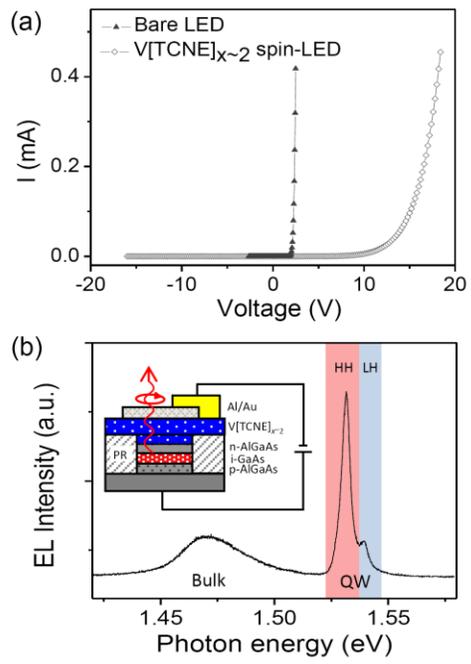

Fig. 3.

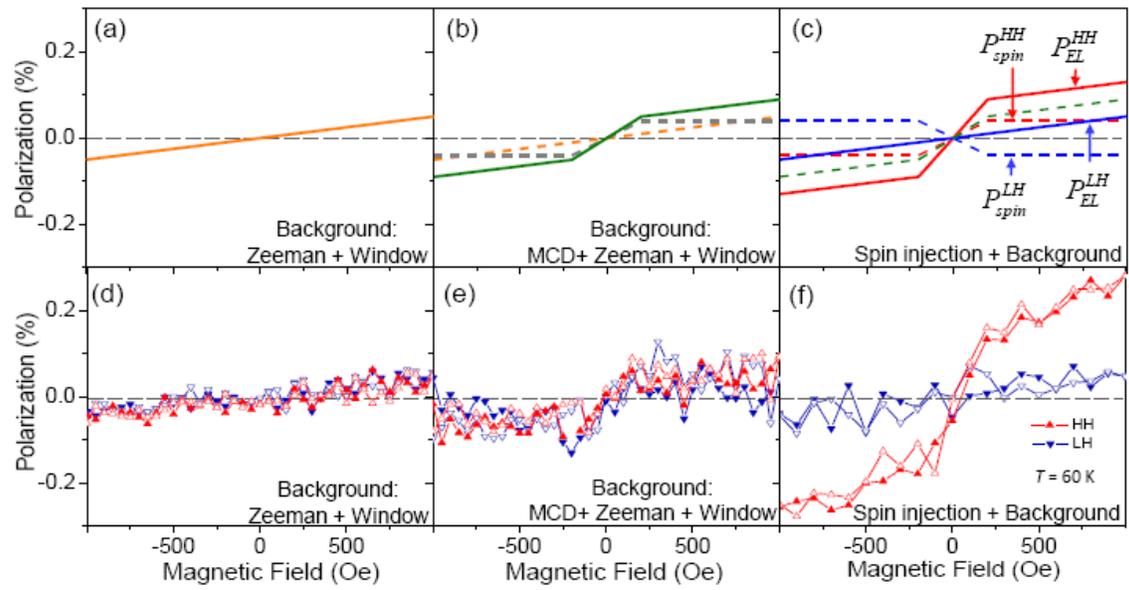

Fig. 4.

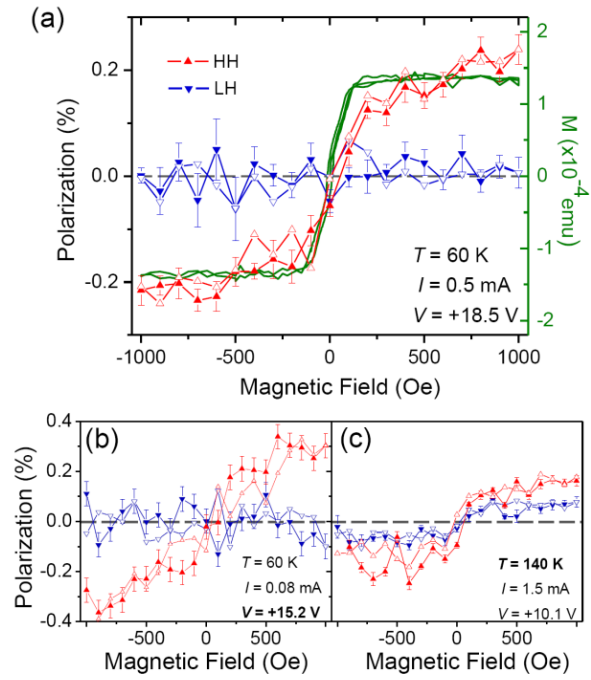